\documentclass[a4paper,11pt]{article}
\pdfoutput=1 

\usepackage{jcappub} 

\usepackage[T1]{fontenc} 
\usepackage{comment}
\usepackage{xspace}
\usepackage[dvipsnames]{xcolor}

\usepackage{hyperref}

\newcommand{\beq}{\begin{equation}}
\newcommand{\eeq}{\end{equation}}

\title{\boldmath Identification of low energy neutral and charged cosmic ray events in large wide field observatories}

\author[a,b]{L.~Apolin\'{a}rio}
\author[a,b]{P.~Assis}
\author[b]{P.~Brogueira}
\author[a,b,1]{R.~Concei\c{c}\~ao\note{Corresponding author. e-mail: ruben@lip.pt}}
\author[a,b]{P.~J.~Costa}
\author[a,c]{G.~La~Mura}
\author[a,b]{M.~Pimenta}
\author[a,b]{B.~Tom\'e}

\affiliation[a]{Laboratório de Instrumentação e Física Experimental de Partículas (LIP),\\ Av. Prof. Gama Pinto, 2, P-1649-003 Lisbon, Portugal}
\affiliation[b]{Departamento de F\'isica, Instituto Superior T\'{e}cnico (IST),\\ Universidade de Lisboa, Av. Rovisco Pais 1, 1049-001, Lisbon, Portugal}
\affiliation[c]{INAF - Osservatorio Astronomico di Cagliari, Selargius, Italy}

\abstract{The lower energy thresholds of large wide-field gamma-ray observatories are often determined by their capability to deal with the very low-energy cosmic ray background. In fact, in observatories with areas of tens or hundreds of thousands of square meters, the number of background events generated by the superposition of random, very low energy cosmic rays is huge and may exceed by far the possible signal events. 
In this article, we argue that a  trigger strategy based on pattern recognition of the shower front can significantly reject the background, keeping a good efficiency and a good angular accuracy (few square degrees) for gamma rays with energies as low as tens of GeV. In this way, alerts can be followed or emitted within time lapses of the order of the second, enabling wide-field gamma-ray observatories to better contribute to global multi-messenger networks of astrophysical observatories.}

\begin{document}
\maketitle
\flushbottom

\section{Introduction}
\label{sec:intro}

The observation of transients in the sky was always a source of great interest, even of astonishment, for the general public and a privileged object of study for scientists.
The sky is, in this way, constantly scrutinized by a large number of detectors exploring all possible particle messengers and, in particular, photons within a very wide range of energies. 
At the gamma-ray energy region, satellite experiments cover the sky up to energies of a few tens of GeV. 
Above these energies, ground-based observatories assume these observations. These include Imaging Atmospheric Cherenkov Telescopes (IACT) and Extensive Air Shower (EAS) arrays.

IACTs~\cite{MAGIC_Crab,HESS,CTA} exhibit favorable attributes such as low energy thresholds (within the range of a few tens of GeV) and exceptional angular resolution (a few arcminutes). However, they also come with limitations, including a relatively narrow field of view (typically on the order of a few degrees) and lower duty cycles (around 10 - 15\%).

Conversely, contemporary EAS arrays~\cite{HAWC_Crab,LHAASO_WCDA_Crab} feature significantly broader fields of view (measured in steradians), along with nearly complete duty cycles, close to 100\%. Yet, they are accompanied by certain drawbacks, notably diminished angular resolution (around 1-2 degrees at lower energies) and higher energy thresholds (in the range of a few hundreds of GeV).

In fact, in an array of Water Cherenkov Detectors (WCD), as described below and in an interval of a few hundred nanoseconds, the expected level of signal and background rates (see section ~\ref{sec:rates}) will imply several tens of hit stations and data flow rates
that could reach unsustainable levels of gigabytes per second. 
Such high rates present a scenario untenable for a large-scale gamma-ray array experiment. 
To effectively handle this, a low-energy trigger system must achieve background rejection factors within the range of $10^3 - 10^5$. 

However, a trigger/analysis strategy that relies solely on the number of hit stations, would require large multiplicity thresholds, which would consequently increase the energy thresholds for shower events.  While more sophisticated methods tailored to the specific signal characteristics could help achieve lower energy thresholds without overwhelming the DAQ, relying exclusively on simple criteria may not be effective.

The strategy explored in this article is based on the recognition of the space-time correlations of the particles in the shower front arriving at one WCD ground array. In this analysis, the shower front was treated as a planar structure.

The detector under consideration occupies an area of $80\,000\,{\rm m^2}$ and is positioned at $5200\,$m above sea level. The array  configuration, with a circular shape,  was emulated by a two-dimensional histogram with cells with an area of $\sim 12\,{\rm m^2}$; a correction was applied to account for  a Fill Factor (FF) of $80\%$. The chosen array layout corresponds to the baseline layout for the future Southern Wide-field Gamma-ray Observatory (SWGO)~\cite{SWGO}. 

Station triggering was governed by a probability function that effectively reproduces the trigger rates observed by the High-Altitude Water Cherenkov (HAWC) observatory, as detailed in Section \ref{sec:rates}. A $200\,$ns sliding window is used to find the shower plane in between all the noise. Experiments like the one considered here typically use a time window of $150$-$300\,$ns~\cite{HAWCtrigger1, HAWCtrigger2,LHAASOtrigger1, LHAASOtrigger2}. This range enables sampling a substantial portion of the shower front while minimizing spurious noise from unrelated sources. In this study, we selected $200\,$ns as a reasonable window (following this work~\cite{LATTES}). This is a conservative approach given that larger time windows are generally required to capture more inclined showers fully across the array.

In essence, the algorithm designed to trigger individual shower events can be succinctly summarized as follows:

 \begin{itemize}    
     \item 
     The arrival time, $t_i^0$, of the shower plane in each triggered cell was defined as the time of the first particle detected in the detector cell while accounting for the detector time resolution; 
     \item 
     The event is characterized by the number, locations and $t_i^0$s of its triggered cells;
     \item 
     The  planes defined by the combination of the time and space coordinates of any triplet of triggered  cells, were characterized by their unitary normal vectors;
     \item
     The direction of the normal vectors were clustered using a QCD jet inspired algorithm;
     \item 
     The direction of the clustered jet vector was taken as the incoming particle direction. 
     \item 
     To complete the shower plane definition, the barycenter of all cells was taken as a reference point in the $\rm (X,Y)$ plane.
     \item 
     For all planes defined by a triplet of cells, the $\rm Z\equiv c t_i^0$ coordinates, where $c$ is the speed of light in vacuum, of the interception with the vertical line crossing the reference point were accumulated into a histogram.
     \item 
     The plane was taken as being centered at the $\rm Z$ of the bin with the most entries, with a width defined by the distance from the peak to the bin with $\rm 10\%$ of its entries.
      \item
    The directions of the selected clustered jets are used then to update a cumulative lookup table of sky  directions.
       \item
    Whenever there will be an accumulation of candidates in one of the cells of the lookup table, an alert will be emitted.
 \end{itemize}

The structure of this article is outlined as follows: In Section~\ref{sec:rates}, we elaborate on the simulation sets, providing a description thereof, while also estimating the anticipated background and signal rates. Section~\ref{sec:trigger} is dedicated to the development of the trigger concept itself, accompanied by an assessment of its performance. Moving to Section~\ref{sec:implementation}, we explore a plausible implementation of this trigger through an FPGA-based system coupled with a microprocessor. Additionally, we briefly present its potential to generate multiple tiers of rapid alerts. Lastly, in Section~\ref{sec:conclusions}, the use of this trigger strategy in future ground array gamma-ray observatories is discussed.

\section{Signal, background and simulations}
\label{sec:rates}

Charged cosmic rays can be studied via Extensive Air Showers (EAS) for energies surpassing a certain threshold while concurrently constituting a formidable background source for high-energy gamma rays. This background emerges from low-energy showers (with primary energies roughly below a few tens of GeV), frequently resulting in the generation of isolated muons at ground level, commonly referred to as \emph{atmospheric muons}.

The rate of the atmospheric particles at 5200 m a.s.l. was estimated, considering a WCD with an area of $12\,{\rm m^2}$,
to be of the order of  $20-30\times 10^3 \,{\rm s^{-1}}$~\cite{WH_private}. Similar rates were obtained by the authors using CORSIKA~\cite{CORSIKA} simulations. Here the simulated showers were sampled at an altitude of $5200\,$m a.s.l. and considering the magnetic field for the geographic location of the ALMA site~\cite{ALMA}. The simulation framework of the LATTES experiment~\cite{LATTES} -- based on the Geant4 toolkit~\cite{Geant4_2016,agostinelli2003geant4,Geant4_2006} --  was used to simulate the detector response\footnote{The station trigger condition requires a minimum of 10 photoelectrons.}. From this simulation exercise, it was verified that the station trigger efficiency was roughly $50\%$ for an electromagnetic energy deposition of $15\,$MeV, similar to what is quoted in~\cite{WH_private} for the nominal trigger efficiency curve of a detector unit, with a scale factor for the photon detection efficiency of one. 
The rate of the possible dark noise and after-pulses able to trigger the cell (detector) in a $200\,$ns time window was considered to be small and within the uncertainty of the atmospheric muons estimation.

The probability that one shower secondary particle triggers a cell is, following reference~\cite{WH_private}, defined as:

\begin{equation}
    1- \exp(- F \,  k \, E),
\label{eq:Probability}
\end{equation}

where $k$ was set to $3.86105 \times 10^{-2} \, {\rm MeV^{-1}}$, $F$ is a scale factor and $E$ the particle energy in MeV.
The parameter $F = 1$ describes the probability to detect at least one photoelectron (p.e) in a station of the central array of the HAWC experiment~\cite{HAWC_Crab}.
In this work, conservatively, $F$ was set to one but it would be possible, if enough computing power is available (see section \ref{sec:implementation}), to increase $F$ and thus lower the energy threshold, at least during a given time interval after external or internal alerts.

\begin{figure}[!t]
  \centering
  \includegraphics[width=0.45\textwidth]{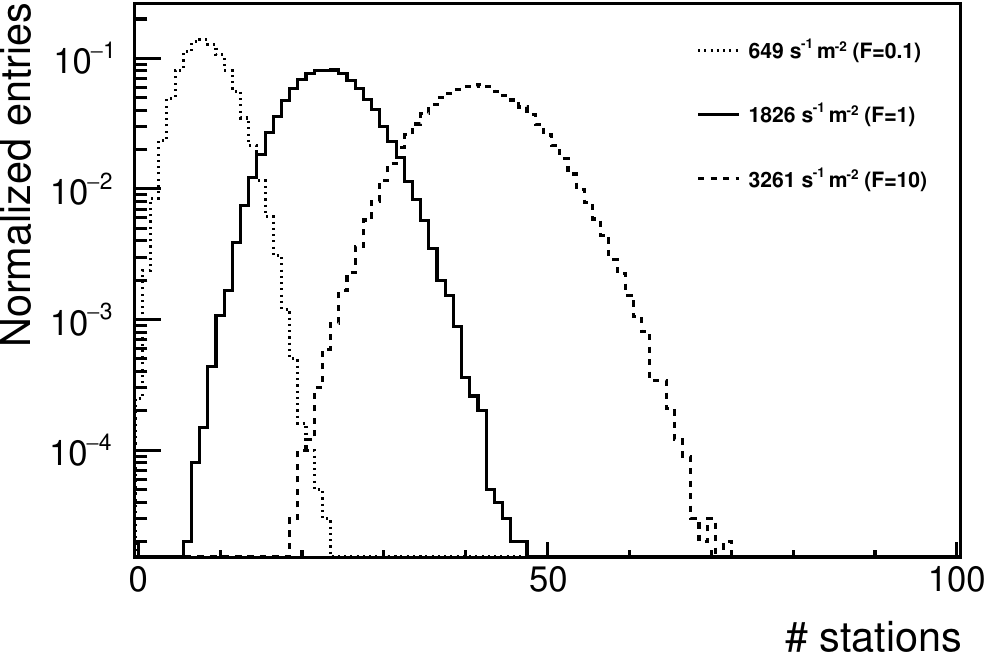}
  \includegraphics[width=0.45\textwidth]{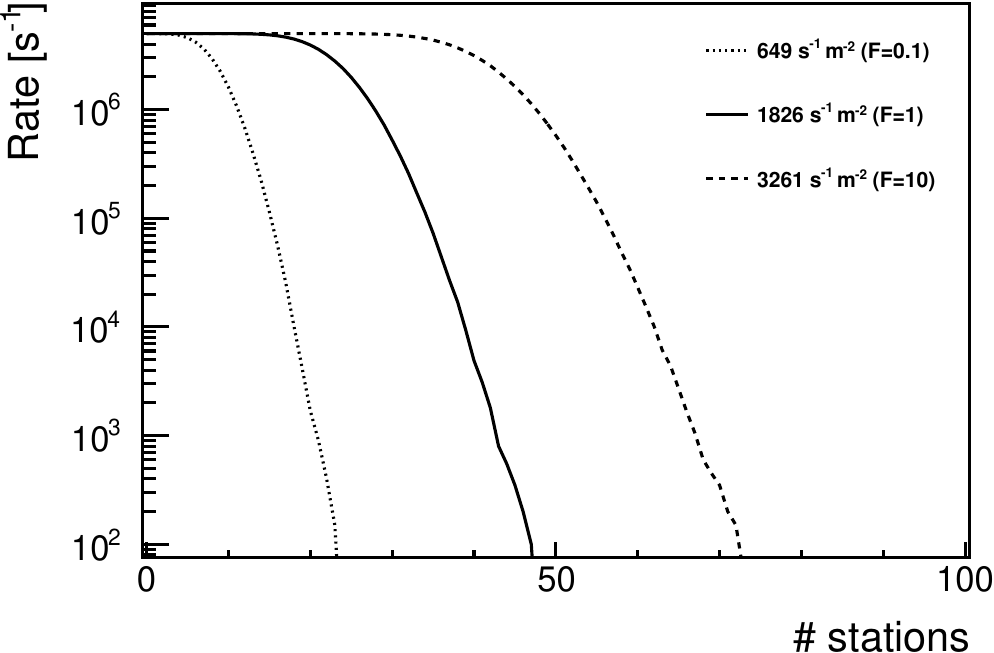}
  \caption{\label{fig:nstations_back} (left) Number of active stations in 200\,ns for different cosmic-ray particle detection rates at 5200\,m altitude, corresponding to scale factors $F = 0.1, 1$ and $10$. (right) The corresponding rate of triggers on the cosmic-ray background as a function of the minimum number of active stations.
}
\end{figure}

The distributions of the number of stations triggered by atmospheric muons in a time window of $200\,$ns are shown in Fig.~\ref{fig:nstations_back} (left) for scale factors $F$ of $0.1$, $1$ and $10$. For a scale factor of one, the mean number of background stations is $23$.
The right plot of Fig.~\ref{fig:nstations_back} shows the corresponding variation of the trigger rate with an increasing threshold on the number of active stations.  For a scale factor of one and a minimum number of active stations lower than $20$, the trigger is saturated, corresponding to a trigger rate of $5 \times 10^6\,{\rm s^{-1}}$.

The integrated flux of charged cosmic rays, with energies higher than a given energy $E$ can be approximated by,

\begin{equation}
    \Phi (>E)  \simeq  10^{4} \left(\frac{E}{{\rm GeV}}\right)^{-1.7} \mathrm{m^{ -2} \, s^{ -1} \, sr^{-1}}
\end{equation}

where $E$ is the energy of the primary cosmic ray.
Assuming a field of view of $2\,$sr, the rate of charged cosmic rays hitting the array with an energy greater than $100\,(200)\,$GeV is about $600\,(200)\times 10^3\,{\rm s^{-1}}$.


The gamma and proton-induced shower simulations used in this work were produced using CORSIKA (version 7.5600)~\cite{CORSIKA}, taking FLUKA~\cite{fluka,fluka2} and QGSJet-II.04~\cite{QGS} as the low- and high-energy hadronic interaction model, respectively.
The simulated events were sampled at an altitude of $5.2\,$km a.s.l., with the shower core randomized uniformly within the array area.
The energies were chosen to be between $10$ and $250\,$GeV for gamma and proton-induced showers. The zenith angle $\theta$  was set to $10^\circ$ or $30^\circ$, and the azimuthal angle $\phi$ followed a uniform distribution.

Shower particles of a specified energy reaching a given ground cell are considered triggered based on the expression in Eq.~\ref{eq:Probability}. Their arrival times are then smeared according to parameterizations that depend on particle type and energy, derived from a full detector simulation of the station, as detailed in~\cite{Mercedes}.

\begin{figure}[!t]
  \centering
  \includegraphics[width=0.8\textwidth]{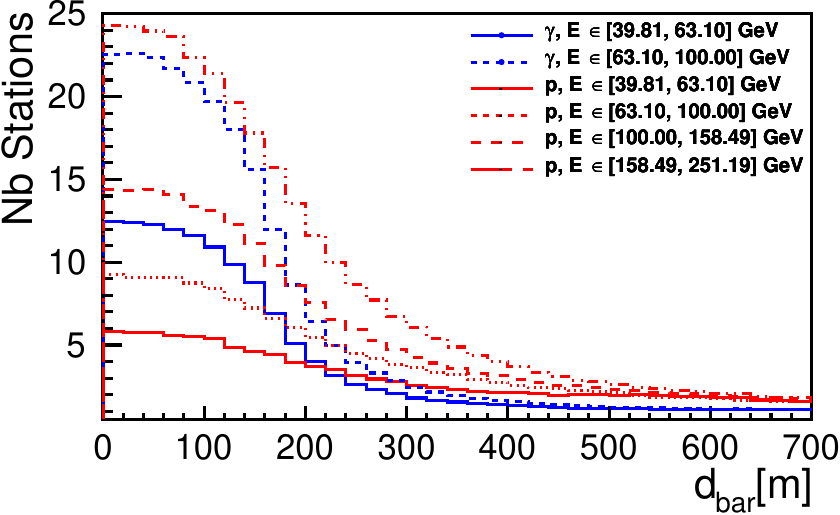}
  \caption{\label{fig:nstations_sig}  Mean number of active stations at 5200\,m  a.s.l  as a function of the distance of the events reconstructed core to the centre of the array ($d_{bar}$), for gamma ray events with a primary energy of about 50 and 80 GeV  (blue lines); the red lines represent the corresponding distributions for charged cosmic rays for primary energies of about 50, 80, 130 and 200 GeV.
}
\end{figure}

The distributions of the mean number of triggered stations in gamma-ray events with primary energies of about $50$ and $80\,$GeV are shown in blue in Fig.~\ref{fig:nstations_sig} as a function of the distance of the events reconstructed core to the centre of the array, $d_{\rm bar}$. For $d_{\rm bar}$ lower than $160\,$m (the array radius), the mean number of triggered stations is between $16$ and $24$ ($80\,$GeV) and between $9$ and $12$ ($50\,$GeV). In the same figure, the equivalent distributions for charged cosmic rays of energies of about $50$, $80$, $130$, and $200\,$GeV are depicted in red. From this figure, it can be concluded that the energies corresponding to the same order of triggered stations are about two times higher in charged cosmic rays than in gamma rays in events where the shower core is lower than $160\,$m.

The background events from atmospheric muons were generated by randomly selecting stations within the array and assigning arrival times uniformly distributed within the specified time window. The average number of background stations for this array configuration was $23$, and for each shower event, this quantity fluctuated according to a Poisson distribution.

It is important to note that there is an overlap between the rates of the generated atmospheric muons and low-energy showers. Given that the primary objective of this work is to demonstrate the potential for lowering the energy threshold through the implementation of a dedicated trigger scheme, a conservative approach was adopted. Specifically, atmospheric muon rates and shower rates were assumed to be independent, providing a cautious estimate for the achievable energy threshold.

\section{ Trigger concept, performances and alerts}
\label{sec:trigger}

\begin{figure}[!t]
  \centering
  \includegraphics[width=0.45\textwidth]{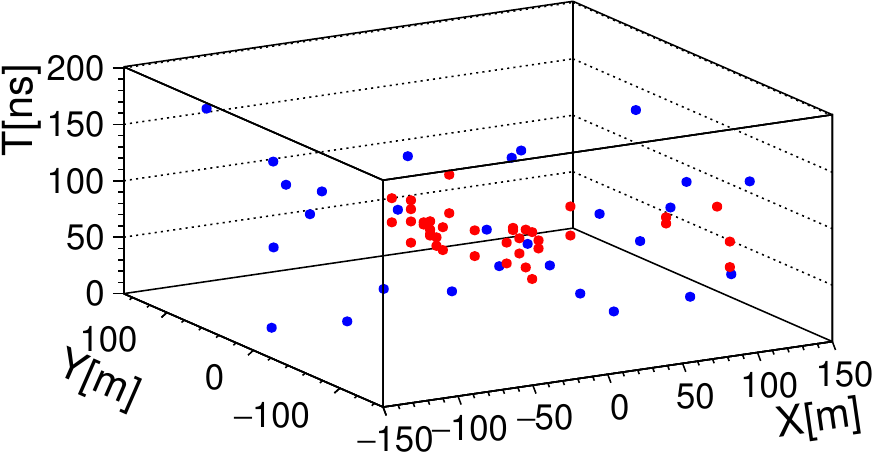}
  \includegraphics[width=0.45\textwidth]{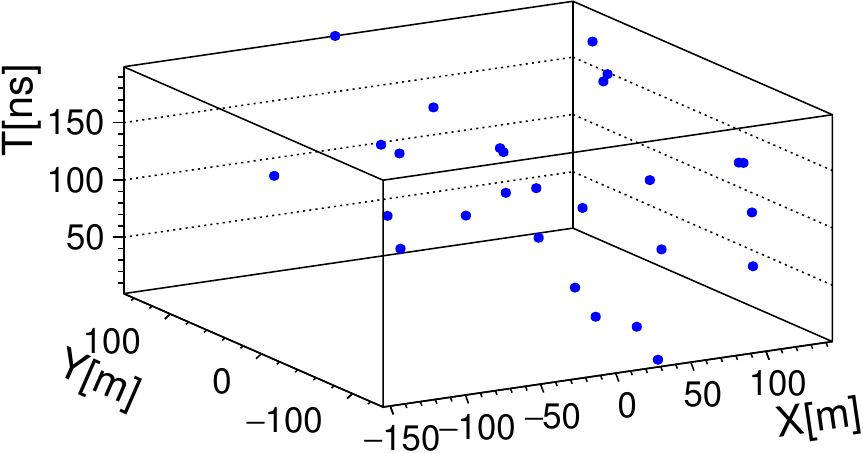}
  \caption{\label{fig:events} (left) A signal event with 19 hit stations (red points) is superposed with 30 background stations (blue points) in a time window of $200\,$ns – the shower plane is trivially seen. (right) A background event with 27 hit stations (blue points)- no evidence for a shower plane is found.
}
\end{figure}

At a given point in time, the particles within a shower front tend to align approximately in a spatial plane. When the shower front reaches the ground, this plane can be reconstructed using the spatial coordinates and arrival times of the impacted stations. In contrast, no such correlation exists for atmospheric muons reaching the ground within the same time window. These distinctions are visually depicted in Fig.~\ref{fig:events}.

Three points in the space define one plane, which can be characterized by its normal vector. The combination of any three triggered stations defines thus a plane and its associated normal vector is designated as $\vec{N}_i$. 
When combining all the $\vec{N}_i$ components in a single shower event, a distinct direction is indicated. Such would not happen for a background event coming from the superposition of tens of low-energy cosmic rays.

\begin{figure}[!t]
  \centering
  \includegraphics[width=0.45\textwidth]{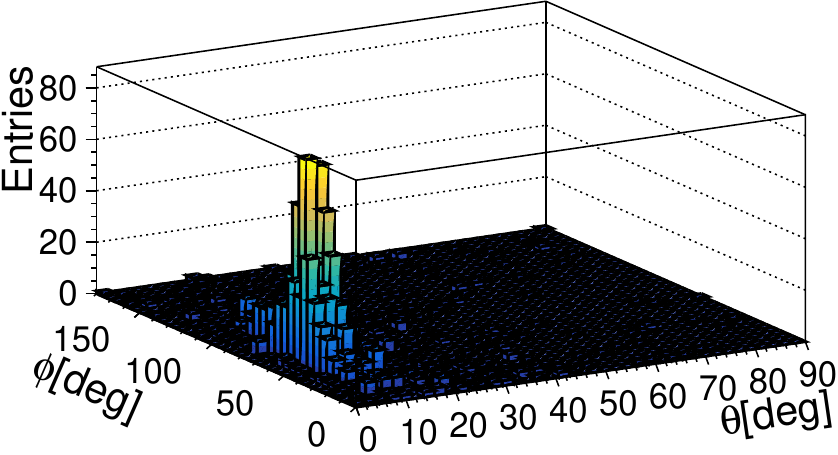}
  \includegraphics[width=0.45\textwidth]{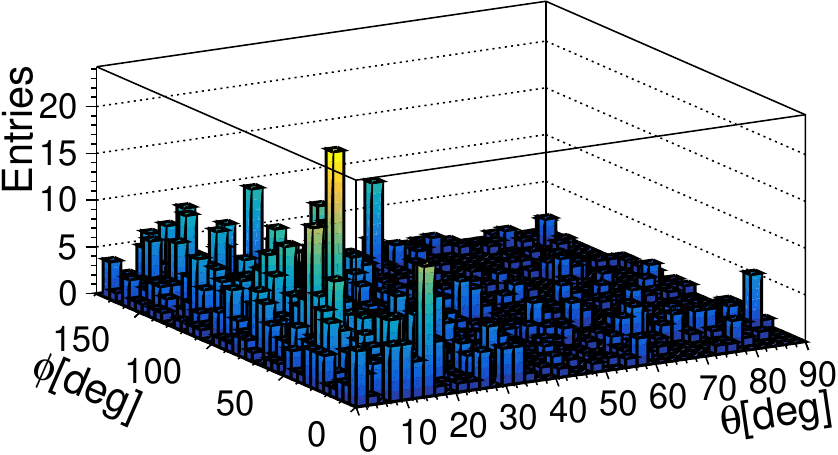}
  \caption{\label{fig:Vectors} Distribution of the directions of  $\vec{N}_i$ vectors in the plane ($\theta \, , \phi$) for a signal event (gamma-induced shower) with an energy of $55\,$GeV, polar angle $10^\circ$ and $23$ hit stations in a $200\,$ns time window (left) and
   for a background event (atmospheric muons) with $19$ hit stations in a $200\,$ns time window (right). 
}
\end{figure}

 As an illustration, in Fig.~\ref{fig:Vectors} are shown the polar, $\theta$, and azimuthal, $\phi$, angles histograms of the directions of $\vec{N}_i$ vectors in a signal (left) and a background event (right). The ($\theta \, , \phi$) histograms were constructed using the partition of the sky semi-hemisphere into 1000 equal-area cells using the prescription detailed in~\cite{hemisphere_partition}.
 A clear peak is seen in the signal event, while no such peak is observed in the background event.
 
\begin{figure}[!t]
  \centering
  \includegraphics[width=0.8\textwidth]{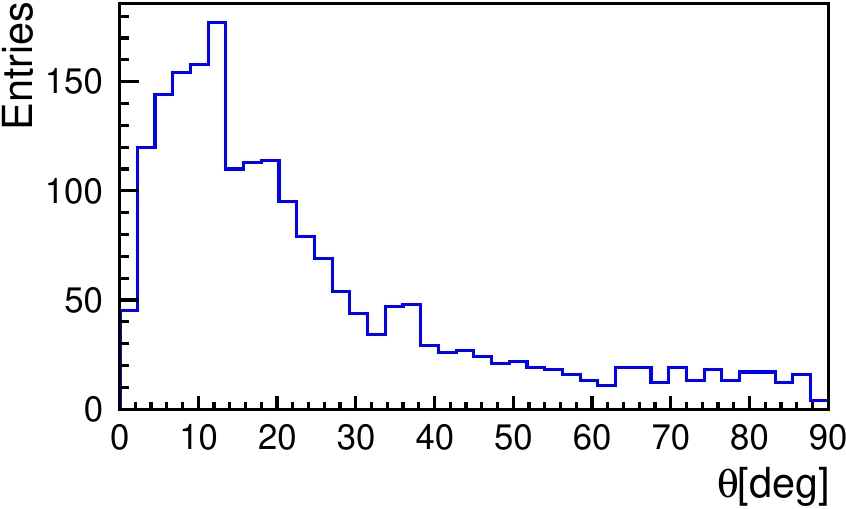}
  \caption{\label{fig:Theta_backg} Distribution of the directions of the $\vec{N}_i$ vectors as a function of the zenith angle $(\theta)$ for a background event with 25 hit stations in a time window of $200\,$ns.
}
\end{figure}

However, in the case of background events, there is a notable accumulation of data points in the vicinity of the vertical direction, as depicted in Figure~\ref{fig:Theta_backg}. It is important to note that this phenomenon is primarily influenced by the length of the time window rather than the number of stations registering hits.

\begin{figure}[!t]
  \centering
  \includegraphics[width=0.8\textwidth]{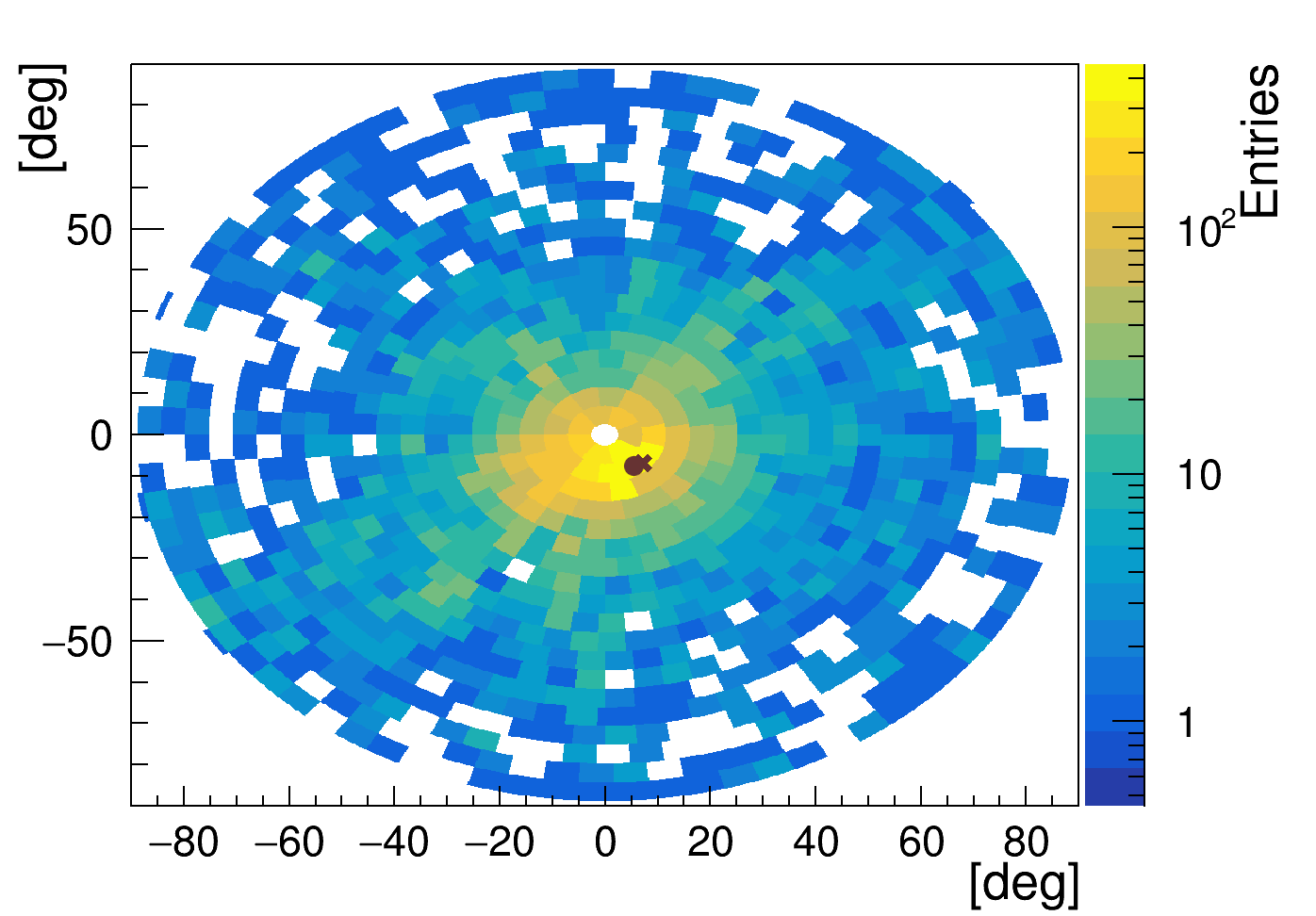}
  \caption{\label{fig:jets_g} Sky semi-hemisphere histogram of the direction vectors arising from the 3-station planes for a single event. This particular event is composed of: one gamma event with energy of $\rm{E_{\gamma}} = 159.8\,$GeV, $\theta = 10^\circ$ and $\phi = -43.2^\circ$; superimposing a mean atmospheric background event. The color scale represents  the size of the combination vectors in each histogram cell. The black cross represents the original direction of the primary gamma and the black spot the reconstructed direction of the most prominent jet (see text for details).  
 }
\end{figure}

\begin{figure}[!t]
  \centering
  \includegraphics[width=0.8\textwidth]{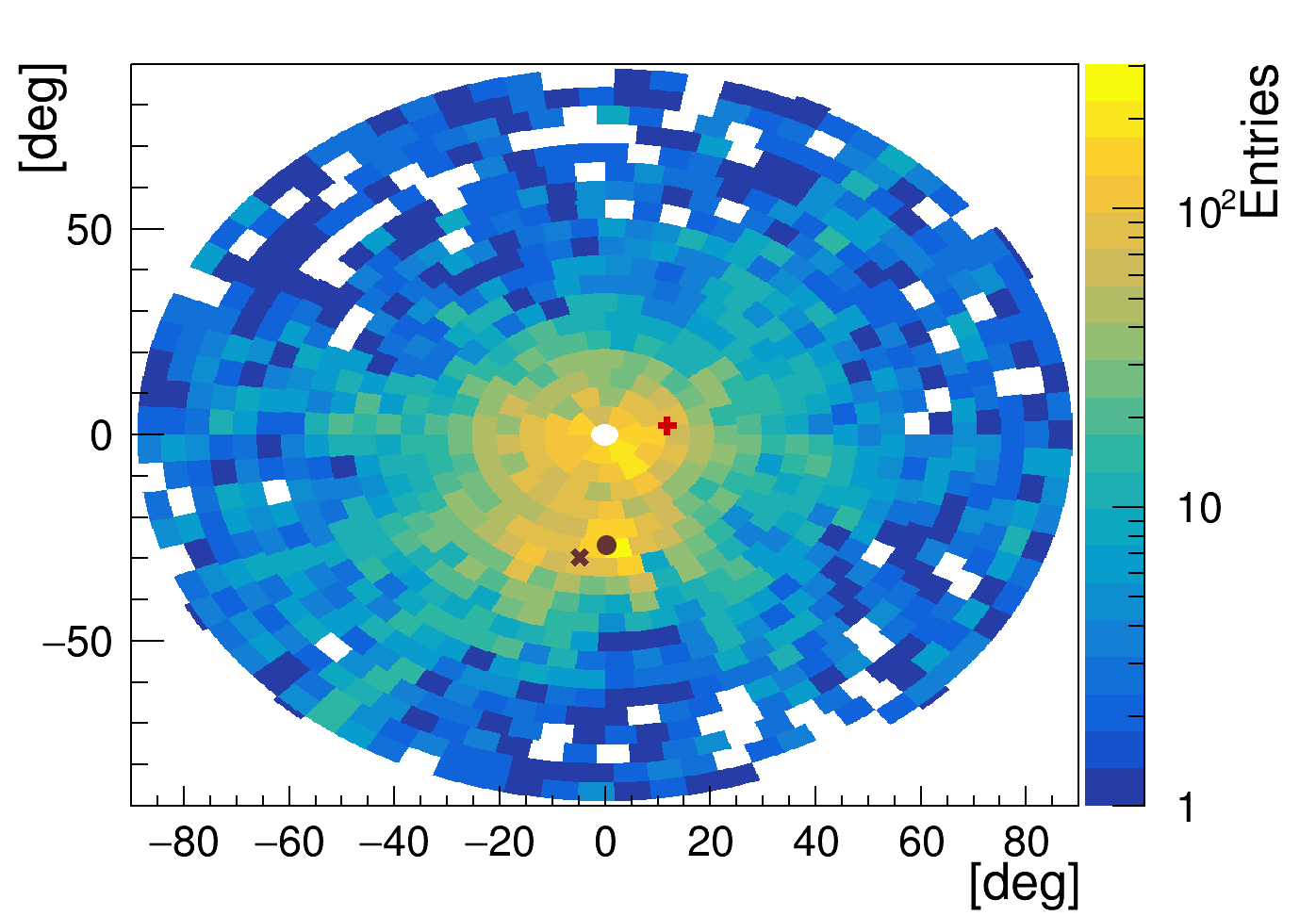}
  \caption{\label{fig:jets_gp} Sky semi-hemisphere histogram of the direction vectors arising from the 3-station plane for a single event. This particular event is composed of: one gamma event with $\rm{E_{\gamma}} = 45.5$ GeV, $\theta = 30^\circ$ and $\phi = -99.1^\circ$;
 one proton with an energy with $\rm{E_{p}} = 155.5$ GeV, $\theta = 12.1^\circ$ and $\phi = 9.9^\circ$; and superimposing a mean atmospheric background event. The color scale represents  the size of the combination vectors in each histogram cell. The black cross represents the original direction of the primary gamma and the black spot the reconstructed direction of the most prominent jet (see text for details).  The original direction of the proton shower is given by a red  cross.}
\end{figure}

This consistent modulation of the $\theta$ angles in the directions of the background vectors $\vec{N}_i$ has been examined as a means to perform a mean subtraction of the expected background, effectively mitigating potential small biases in $\theta$. The results of this operation are illustrated in Figs.~\ref{fig:jets_g} and~\ref{fig:jets_gp}, for instance. The vacant bins within the \emph{sky semi-hemisphere} in these figures result from an overestimation of the background, but this does not significantly affect the much higher signal cells.
 
Once we have arrived at the sky semi-hemisphere plots, created with the 3-station normal plane vectors, $\vec{N}_i$,  and the removal of zenith background modulation, the final step is to find the plane of the shower, by determining its direction and selecting a reference point. The former step was done by applying a QCD-based jet clustering algorithm to the cells content. Such jet algorithms are commonly used in Particle Physics for QCD studies (e.g~\cite{Salam:2010nqg}). To reconstruct the gamma-ray direction from the $\vec{N}_i$ binned in a semi-hemisphere plane, a Cambridge-Aachen algorithm was used~\cite{Dokshitzer:1997in,Wobisch:1998wt}. The method was adapted to reconstruction jets in a sky semi-hemisphere. For that the input coordinate system was changed for collider variable, namely pseudo-rapidity, $\eta$, was considered as a proxy for the polar angle, $\theta$, and the azimuthal angle, $\phi$, taken directly from the sky-coordinate system. Finally, the transverse momentum of each cell is taken as the number of hits in each station. As such, the sky hemisphere coordinate $(\eta, \, \phi) = (0,0)$ was set to match the $(y,\phi) = (0,0)$ in the collider coordinate system. Moreover, the jet radius was set to $R=0.3$ (corresponding to $20^\circ)$. All these steps were performed using the FastJet v.3.3.0 package~\cite{Cacciari:2011ma}.
The algorithm extracts the direction $(\theta, \phi)$ of the gamma-ray shower from the reconstructed leading jet. This is to be understood as the axis of the jet that has higher number of combinations. 

Having obtained the normal of the shower plane, a reference point is still necessary to complete its definition. This is achieved by taking the spatial coordinates of the barycenter, $\rm (X_b, Y_b)$ to define a vertical line in the $\rm (X, Y, Z)$ space and accumulating the values of  $\rm Z$ at which the previously computed 3-station normal planes intersect it, $\rm Z_{int}$, in a histogram. From this histogram the most frequent value of $\rm Z_{int}$ is selected, and an interval is defined around it, such that it contains all values of $\rm Z_{int}$ with a frequency higher or equal than $10\%$ of the maximum frequency. If a station was used to define any of the planes defined by the previously obtained normal and a reference point within the $\rm Z_{int}$ interval, it is then considered as being within the shower plane.

Accounting for 3-station combinations made up exclusively of stations within the shower plane, the event trigger variable is taken as the content of the sky bin corresponding to the previously obtained direction, $(\theta, \phi)$, and is denoted as $C_3^{N}$. 



As an example, we show the polar ($\theta \, , \phi$) histograms of the direction of the jet vectors found in one gamma event (Fig.~\ref{fig:jets_g}) and one event with one gamma and one proton (Fig.~\ref{fig:jets_gp}) superimposed to a mean atmospheric muon background. The colour scale represents the size of the vectors in each histogram cell.

When the value of $C_{3}^N$ exceeds a specified threshold, denoted as $C_{3}^{NT}$, the event is triggered. 

\begin{figure}[!t]
  \centering
  \includegraphics[width=0.8\textwidth]{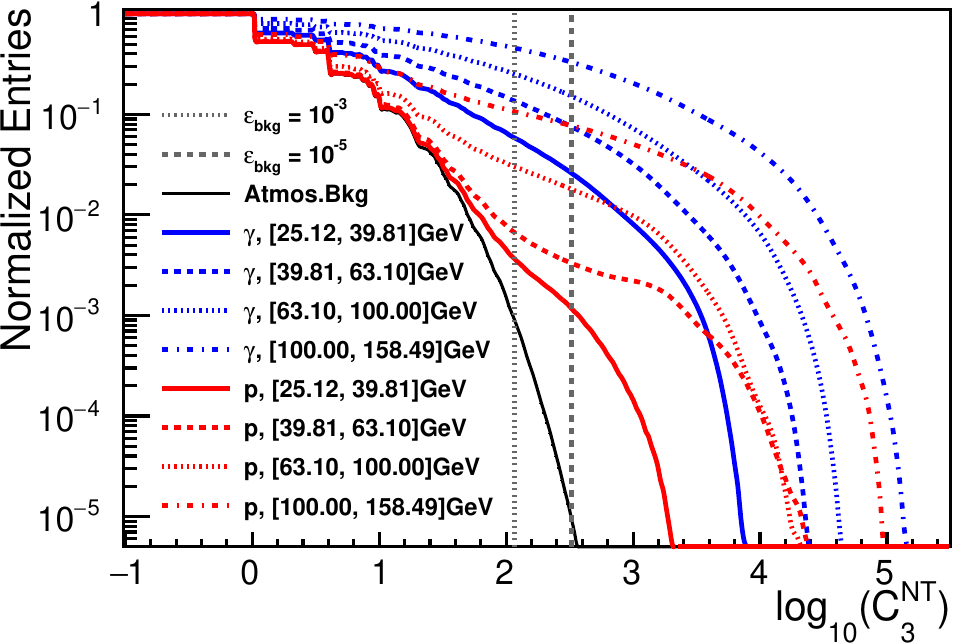}
  \caption{\label{fig:trigger_efficiency} Efficiencies curves as a function of $\log(C_{3}^{NT})$ for atmospheric background events (black line), as well as, the efficiency curves for gammas (blue) and protons (red) for different energies (see legend for details). The vertical dashed (dotted) black line represents the value  of log($C_{3}^{NT}$) needed to ensure atmospheric  background rejection factors of $10^3$ and $10^5$.   
}
\end{figure}

The value of $C_{3}^{NT}$ determines the atmospheric background rejection factors as it is shown in Fig.~\ref{fig:trigger_efficiency}. In this figure, displayed as a function of the value of $\log(C_{3}^N)$, it is shown the efficiency curve for atmospheric background events (black line), as well as the efficiency curves for gammas (blue) and protons (red) with primary energies of about $50$, $80$, $130$ and $200\,$GeV. If $C_{3}^{NT}$ is set to $1000$ (1500) atmospheric background rejection factors  higher than $10^3$ ($10^5$) will be attained.

\begin{figure}[!t]
  \centering
  \includegraphics[width=0.8\textwidth]{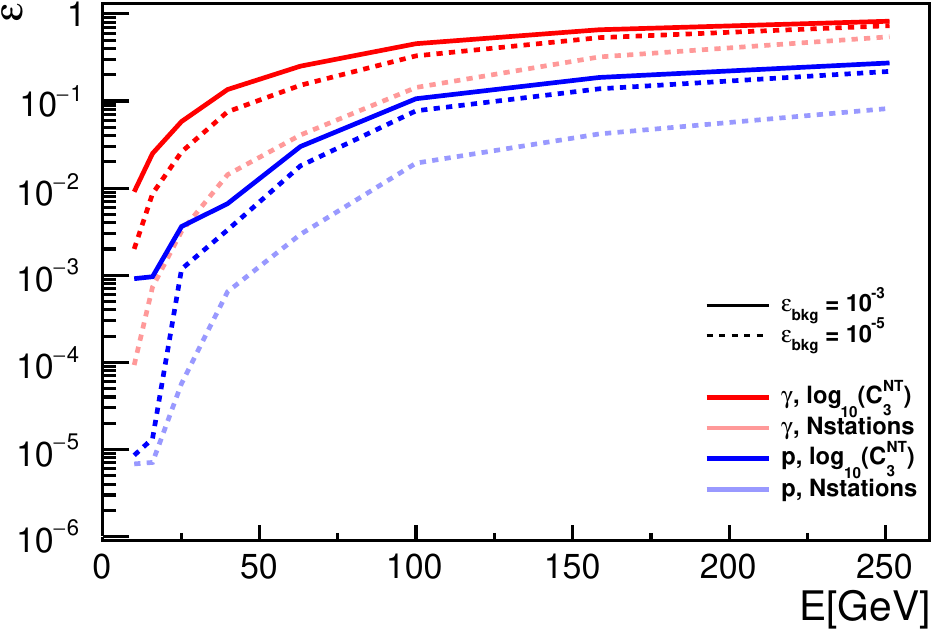}
  \caption{\label{fig:trigger_gh_eficiency} Trigger efficiency as a function of the shower energy for gamma (blue line) and protons (red line), with the trigger threshold set at $C_{3}^{NT}$=117 or $C_{3}^{NT}$=328, corresponding  to atmospheric background rejection factors of respectively $\varepsilon_{\mathrm {bkg}} = 10^{-3}$ and $\varepsilon_{\mathrm {bkg}} = 10^{-5}$.
}
\end{figure}

The trigger efficiencies, with $C_{3}^{NT}$ set to $1000$ and $1500$, are presented in Fig.~\ref{fig:trigger_gh_eficiency} as a function of the shower energy, for both gamma (blue line) and proton (red line) events. The shower events are embedded on the expected atmospheric background. An efficiency rate of $10\%$ corresponds to gamma events with energies around $50\,$GeV and proton events with energies of approximately $150\,$GeV. The proton curve exhibits a flattening at lower energies, which can be attributed to the thresholds set for atmospheric events, at $10^{-3}$ or $10^{-5}$.
 
\begin{figure}[!t]
  \centering
  \includegraphics[width=0.8\textwidth]{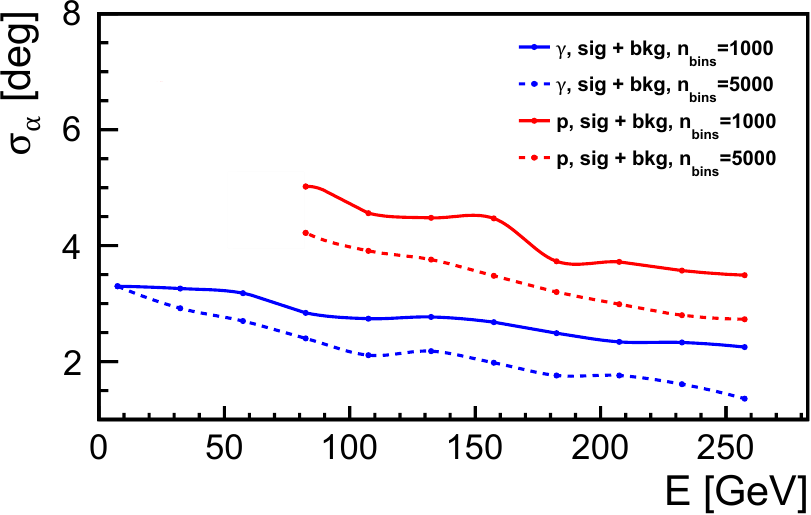}%
  \caption{\label{fig:trigger_angular} Angular accuracy of gamma (blue line) and proton events as function of the energy of the primary gamma, with  $C_{3}^{NT}$=1000. Results are shown for sky semi-hemisphere partitions into 1000 (the default throughout the paper) and 5000 equal-area cells.
}
\end{figure}

On the other hand, the angular accuracy of the tagged events as a function of the shower energy is shown in  Fig.~\ref{fig:trigger_angular}, setting $C_{3}^{NT}$ to $1000$. Even for energies as low as a few tens of GeV, a reasonably high angular accuracy within a few square degrees is assured. 
As expected the resolution improves when the sky semi-hemisphere is partitioned into a greater number of cells ($n_{\rm bins}$). The proton lines are not displayed for energies below $80\,$GeV because the limited number of signal stations at the ground leads to a complete domination of these events by the background, rendering the results meaningless.

At this stage, one might consider whether fitting the shower plane could yield comparable results. Although this approach is certainly feasible, it would involve a complex procedure likely requiring post-event collection processing. In contrast, the method proposed here introduces an algorithm that allows computationally intensive tasks to be executed in parallel using hardware such as FPGAs, as outlined in the next section of the manuscript. Notice that the jet algorithm is applied specifically to identify clusters of events within the sky hemisphere histogram, enabling accurate determination of the triggered shower's direction.

In fact, efficient and timely management of alert reception and issuance has become an imperative aspect in the design of trigger systems for wide-field gamma-ray observatories. When an external alert is received, it becomes necessary to temporarily relax trigger requirements, allowing for the capture of the maximum possible number of events originating from the region in the sky associated with the alert. Conversely, when the observed events from a specific sky region meet predefined criteria in terms of their number or characteristics, alerts should be disseminated to the global network of Astrophysics Observatories, all while maintaining a stringent standard to minimize the occurrence of false alerts.

\begin{figure}[!t]
  \centering
 \includegraphics[width=0.8\textwidth]{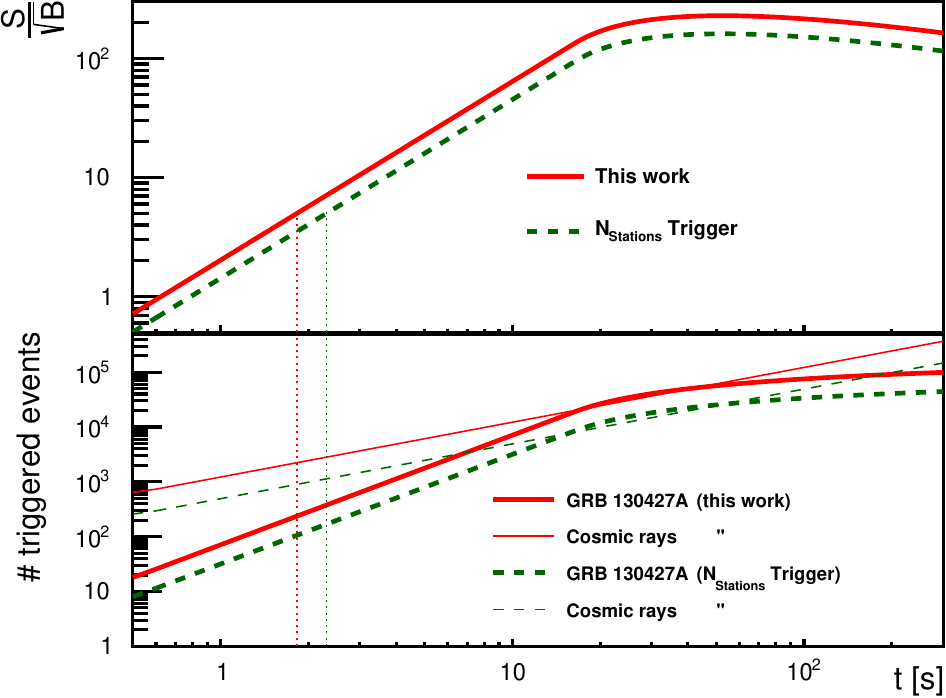}
\caption{\label{fig:alert_example}(top) $\mathrm {S}/\sqrt{B}$ as a function of integrated time for the signal from {GRB 130427A}, for a trigger threshold corresponding to an atmospheric background rejection $\varepsilon_{\mathrm {bkg}} = 10^{-3}$. The curves represent the trigger based on the number of triggered stations (green dashed) and the trigger described in this work (red solid). (bottom) Integrated number of triggered events from cosmic rays and the GRB as a function of time for the two trigger strategies. The vertical dashed lines indicate the time at which $\mathrm {S} = 5 \, \sqrt{B}$.}
\end{figure}

A typical example of the necessity to issue fast alerts is the one connected with the investigation of VHE emission from Gamma-Ray Bursts (GRBs), where the onset of the VHE signal has a critical role in the interpretation of the mechanisms at play. GRBs have now been firmly detected in the VHE domain \cite{MAGIC19,LHAASO2023}, but the frequency and the onset of this spectral component still needs to be understood \cite{LaMura:2021ani}. To illustrate the performance of the trigger strategy discussed above, we can test the characteristics of an event with the spectral and the temporal properties of GRB~130427A, as modeled in~\cite{LaMura:2021hgs}. The expected energy-dependent GRB and cosmic-ray counts accumulated in $80\,000\, \mathrm{m^2}$ are convolved with the corresponding trigger efficiencies. For the cosmic-ray background, only the fraction of events in a solid angle with a $3^\circ$ semi-aperture were considered. This is a fair approximation of photon angular resolution reported in Fig.~\ref{fig:trigger_angular}. In Fig.~\ref{fig:alert_example}, we show the number of accumulated photon counts and cosmic-ray background counts as a function of time for the trigger strategy described in this work and for a trigger based only on the number of active stations. Adopting a threshold of $S \geq 5 \sqrt{B}$ for the excess of signal counts $S$ over background counts $B$ to raise an alert, we see that the requirement would be fulfilled in less than $3\,$s of integration time, with the accumulation of more than 100 counts.

\begin{figure}[!t]
  \centering
 \includegraphics[width=0.8\textwidth]{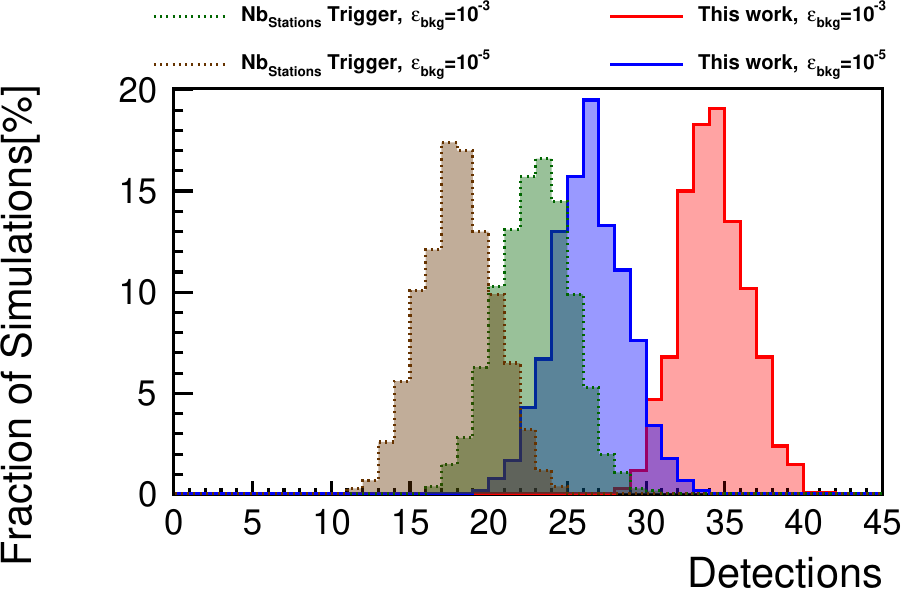}
 \caption{\label{fig:GRB_detections} Fraction of simulations predicting the number of detections shown on the horizontal axis, based on 1000 random redshift distributions for a sample of 140 gamma-ray bursts (GRBs) without measured redshifts, observed by \textit{Fermi}-LAT over 10 years. The plot highlights the improvement in the expected number of detections using the approach developed in this work. The red histogram (\(\varepsilon_{\rm bkg} = 10^{-3}\)) shows that \(\sim 50\%\) of simulations predict more than 34 detections, while the blue histogram (\(\varepsilon_{\rm bkg} = 10^{-5}\)) shows \(\sim 50\%\) predicting more than 26 detections. For comparison, predictions based solely on the number of triggered stations yield lower expected detections: the green/dotted histogram (\(\varepsilon_{\rm bkg} = 10^{-3}\)) has \(\sim 50\%\) of simulations predicting more than 23 detections, and the brown/dotted histogram (\(\varepsilon_{\rm bkg} = 10^{-5}\)) has \(\sim 50\%\) predicting more than 17 detections.
 }
\end{figure}

The impact of this trigger strategy can be further evaluated using an approach similar to the one described in \cite{LaMura:2021hgs} to estimate the probability of detecting VHE emission from GRBs with instruments that are sensitive to radiation down to lower energy limits of $125\,$GeV, $250\,$GeV and $500\,$GeV. In general, the possibility to detect VHE from GRBs depends on three main factors: the GRB brightness (intended as the overall number of emitted photons as function of time); the GRB spectral hardness (affecting the likelihood of producing high energy photons); and the GRB redshift (which controls the VHE photon survival chances to $\gamma\gamma$ pair-production processes). Considering that the ability to access a lower energy spectral window increases the chances of detecting high redshift events, requiring, on the other hand, a larger intrinsic power from the source, we estimated the expected detection rates using the full set of 1000 random redshift distributions computed in \cite{LaMura:2021hgs} for the sample of 140 \textit{Fermi}-LAT detected GRBs without a measured redshift in 10 years of observations \cite{2FLGC}, to derive a statistical sample of possible VHE energy extrapolation of the parent population of GRBs that produced the observed $\gamma$-ray fluxes. Applying the observed temporal and spectral properties and using the simulated redshifts, we tested which fraction of simulations predict the detection of GRBs originating an excess signal obeying the conditions $S \geq 5 \sqrt{B}$ and a signal $S$ with more than 10 detected photons. The results, presented in Fig.~\ref{fig:GRB_detections}, indicate that with the new trigger, the likelihood of detecting signals from GRBs increases by roughly 50\%.

The trigger strategy described here,  devised to efficiently select signal events and handle alerts, must obviously be implemented online, being able to  manage  the rapid increase in the total number of normal vectors, $n_{vec}$, with the increase of the number of hit stations ( $n_{vec} = n_{hit} (n_{hit}-1) (n_{hit-2})/6)$)\footnote{For instances, for $n_{hit} = 10$, $n_{vec} = 120$, while for $n_{hit} = 50$, $n_{vec} = 19\,600$.}.
This is the subject of the following section.
The ability of such trigger strategy to issue alerts is further discussed in section~\ref{sec:conclusions}.

\section{Hardware implementation}
\label{sec:implementation}

The computation of the $\vec{N}_i$ vectors can be done by exploring a parallel computing architecture based on an FPGA-microprocessor system. Indeed, such types of solutions are becoming widely spread. For instance, a considerable effort is being made by Intel\textsuperscript{\textregistered} in this direction.

\begin{figure*}[!t]
  \centering
  \includegraphics[width=0.9\textwidth]{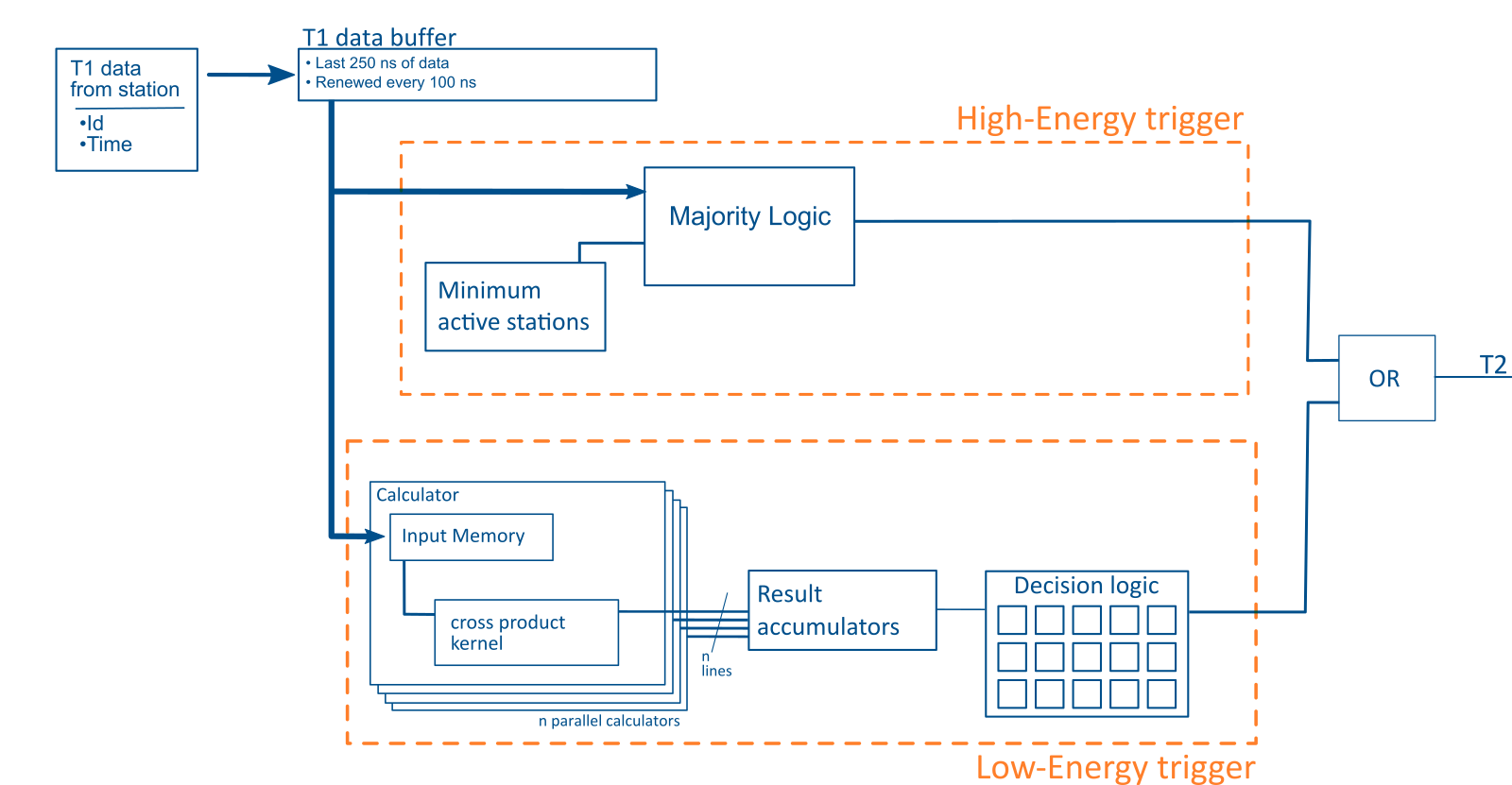}
  \caption{\label{fig:trigscheme} Simplified scheme of a possible architecture of the trigger system. The low-energy trigger final decision is computed in a decision logic array that can be implemented as a CPU farm.}
\end{figure*}

In Fig.~\ref{fig:trigscheme}, a possible architecture of the trigger system is shown. Each station is continuously taking data, and a station-level trigger, T1, is generated each time the station registers a signal above the electronic noise. 
The information of the T1 is transmitted to the Central Unit Network Trigger and kept in a pipeline (station ID, space and time coordinates). Each $100\,\mathrm{ns}$, the various T1 that have arrived in the last $200\,\mathrm{ns}$  are transferred to a buffer memory. If the number of T1 transferred to the buffer is higher than a given threshold (of about 70), a high multiplicity trigger is generated, activating the T2. Otherwise, the trigger will be handled by the \emph{Low Energy Trigger} branch.

The Low Energy Trigger branch is composed of a large set of $n$ trigger modules to compute, in parallel,  the normal vectors $\vec{N}_i$ and the $Z_i$ of each event,  and also a set of CPUs to handle, in parallel, the jet clusterization of computed  $\vec{N}_i$ vectors and the mean of the computed $Z_i$ in each candidate event and, finally, to take a  decision whether the event should be accepted or not.  

Each trigger module is formed by an input memory, which is a mirror of the input buffer, and a calculator. All the modules' input memories are loaded in parallel. The calculator of each module computes the normal to the plane defined by a particular set of three T1 stations. These stations correspond to specific hard-coded positions of the module input memory.
If the number of stations in the buffer is 70, there are $\sim 54\,700$ different combinations of 3 stations to be processed by the $n$ modules, which are implemented using FPGAs.  

The direction angles of the $\vec{N}_i$ vectors of one event, computed by the calculators, are accumulated in a histogram kept in a memory array, which the cluster of CPUs may access. Each CPU will sequentially take care of one event, running the clusterization algorithm and taking the trigger final decision on that event.

To gain insight into the resource requirements for implementing the calculators of the trigger modules using an FPGA system, we implemented a basic algorithm in the Quartus suite by Intel using Verilog. This was done on a mid-low range FPGA and compiled without any optimizations. The resource allocation for this implementation involved 112 logic elements (ALMs) and 12 DSP blocks, all implemented using asynchronous logic. It is worth noting that the desired high level of parallelization can be readily achieved with high-range FPGAs.

Additionally, Intel is going to release the AGILEX family dedicated to intensive computations. It is then possible to implement in one of these FPGAs around $2\,000$ calculators and, using a system of 10 Stratix FPGAs,  up to  $80\,000$  external products can be calculated in  $10\,\mathrm{ns}$. 

On the other hand, suppressing most empty histogram cells,  the mean execution time of the jet sequential clusterization algorithm in one event is, considering a  $2403.0\,$MHz CPU, about  $10\,\mu{s}$. 
In this way, a cluster of a few tens of CPUs may handle trigger rates of a few $10^6\,{\rm s^{-1}}$ and thus comply, at a reasonable cost, with the expected signal rates of low energy events, which deposited a total electromagnetic energy at the ground equivalent to a 100 GeV proton.

\section{ Discussion and Conclusions}
\label{sec:conclusions}

In the context of Gamma Ray Large Wide-Field Observatories, maintaining low energy thresholds is pivotal in gathering robust statistical data. This significance is underscored by its application in two crucial scenarios:
 \begin{itemize}
\item Transient phenomena, like the VHE emission from GRBs and flares, following and issuing alerts;
 \item  Extended sources -- such as the Fermi bubbles -- and diffuse gamma emission, enlarging the energy range of the satellite experiments.
\end{itemize}

This study suggests that upcoming wide-field gamma-ray observatories, such as the forthcoming Southern Wide-field Gamma-ray Observatory (SWGO) \cite{SWGO}, have the potential to operate with remarkably low energy thresholds, reaching levels as low as tens of GeV. This advancement could result in substantial background rejection factors ranging from $10^3$ to $10^5$ and achieve mean angular resolutions, even at the trigger level, at approximately $3$ degrees. Such capabilities would enable round-the-clock alert issuance and monitoring throughout the year, effectively functioning as a monitoring observatory and trigger instrument for the Cherenkov Telescope Array (CTA), as well as gravitational waves and neutrino observatory.

The hardware implementation proposed (as detailed in Section \ref{sec:implementation}) should be regarded as an initial, foundational step to assess the feasibility of a cost-effective implementation using current technology. 

For future developments in trigger systems, it's imperative to closely monitor advancements, particularly at facilities like the Large Hadron Collider, where new triggering systems are adopting similar approaches but at significantly higher levels of complexity, especially in terms of online tracking reconstruction. These algorithms are also at the heart of fields such as computer vision, automated robot navigation, and autonomous vehicle guidance systems, all of which are evolving rapidly.

\acknowledgments

The authors wish to express their appreciation for the financial support provided for this work by FCT - Fundação para a Ciência e a Tecnologia, I.P., through project PTDC/FIS-PAR/4300/2020 and EXPL/FIS-PAR/0905/2021. P.~C. is grateful for the financial support by FCT under UI/BD/153576/2022. L.~A. is grateful for the financial support by FCT under 2021.03209.CEECIND. G.~L.~M. is grateful for the financial support by 1.05.01.01 Fundamental research in Astrophysics under contract n. INAF-OAC-12/2023.

\bibliographystyle{JHEP}
\bibliography{References.bib}

\end{document}